\documentclass[aps,superscriptaddress,twocolumn,a4paper,reprint]{revtex4}
\usepackage[colorlinks=true, pdfstartview=FitV, linkcolor=blue, citecolor=red, urlcolor=black]{hyperref}
\usepackage{graphicx}
\usepackage[all]{xy}
\usepackage{amsmath}
\usepackage{amssymb}
\usepackage{tensor}
\usepackage{orcidlink}
\newcommand{\be}{\begin{equation}}
\newcommand{\ee}{\end{equation}}
\newcommand{\ben}{\begin{eqnarray}}
\newcommand{\een}{\end{eqnarray}}
\newcommand{\bes}{\begin{subequations}}
\newcommand{\ees}{\end{subequations}}
\def\bal#1\eal{\begin{align}#1\end{align}}

\newcommand{\nn}{\nonumber\\}
\newcommand{\bfi}{\begin{figure}}
\newcommand{\efi}{\end{figure}}
\newcommand{\bc}{\begin{center}}
\newcommand{\ec}{\end{center}}

\newcommand{\arctanh}{\mbox{arctanh}}

\newcommand{\sgn}{\mbox{sgn}}

\newcommand{\sech}{\mbox{sech}}
\newcommand{\arcsinh}{\mbox{arcsinh}}

\newcommand{\LL}{{\cal L}}

\newcommand{\n}{\nabla}

\begin{document}

\title{Flat and bent branes in Born-Infeld-like scalar field models}
\author{I. Andrade\,\orcidlink{0000-0002-9790-684X}}
        \email[]{andradesigor0@gmail.com}\affiliation{N\'ucleo de Forma\c{c}\~ao de Docentes, Universidade Federal de Pernambuco, 55014-900, Caruaru, PE, Brazil}
\author{M.A. Marques\,\orcidlink{0000-0001-7022-5502}}
        \email[]{marques@cbiotec.ufpb.br}\affiliation{Departamento de Biotecnologia, Universidade Federal da Para\'\i ba, 58051-900 Jo\~ao Pessoa, PB, Brazil}
\author{R. Menezes\,\orcidlink{0000-0002-9586-4308}}
     \email[]{rmenezes@dcx.ufpb.br}\affiliation{Departamento de Ci\^encias Exatas, Universidade Federal
da Para\'{\i}ba, 58297-000 Rio Tinto, PB, Brazil}


\begin{abstract}
In this work, we investigate the presence of thick branes modeled by a single scalar field with Born-Infeld-like dynamics. We consider the 4-dimensional metric being Minkowski, de Sitter or anti-de Sitter. We obtain the field equations and the conditions to get a first order formalism compatible with them. To illustrate our procedure, some specific models are presented. They support localized warp factor and have their properties controlled by the 4-dimensional cosmological constant. In particular, a hybrid brane may arise, with a thick or thin profile depending on the extra dimension being inside or outside a compact space.

\end{abstract} 

\maketitle

\section{Introduction}
Higher dimensional theories supporting braneworlds attracted much interest as they provide a tentative solution for the hierarchy problem \cite{AH,RS1,RS2,GW}. In particular, in Ref.~\cite{RS2}, it was proposed a brane model with a thin profile. Later, scalar fields were included to modeling the extra dimension, giving rise to the so-called thick brane \cite{cvetic,ST,csaki,dewolfe,gremm1,gremm2}. In particular, in Ref.~\cite{dewolfe}, the scalar field was introduced to obtain a first-order framework compatible with the equations that describe the brane.

Over the years, several papers have dealt with thick branes \cite{t1,t2,t3,t4,t6,t7,t8,t9,t10,t11,t12,t13,t14,t15,t16,t17}. For instance, in Ref.~\cite{t3}, a model with two scalar fields was investigated, leading to the Bloch brane, which may engender internal structure. In Ref.~\cite{t8}, the localization of fermions was studied and the appearance of resonances was related to their internal structures. Symmetric and asymmetric branes with a hybrid character determined by the width of compact scalar field solutions were proposed in Ref.~\cite{t13,t14}. A comprehensive review about thick branes can be found in Ref.~\cite{t9}.

Studies of branes also considered the possibility of introducing a 4-dimensional cosmological constant in the models, leading to \emph{bent} branes \cite{kaloper}. In this situation, the investigation becomes quite harder, as the presence of this constant adds a term in the second order differential equations depending explicitly on the warp function. Even so, some solutions were found; see, for instance, Refs.~\cite{slatyer,kobayashi,sasakura1,wang}. In order to get first order equations, a method was proposed in Ref.~\cite{dewolfe} and its features were further explored in Refs.~\cite{gremm1,sasakura2,freedman}. Later, in Ref.~\cite{bazeiabent}, the authors introduced a new first order framework, for both de Sitter and anti-de Sitter geometry, and presented solutions considering specific functions of the scalar field that allows for the constraints to be solved in a simpler manner. Other papers which have used this approach include Refs.~\cite{britobent,britobent2}.

In the thick braneworld scenario, the scalar field usually appears with a kinklike profile, from a Lagrangian density that consists of the difference between the dynamical and potential terms. In the $(1,1)$-dimensional scenario, these structures may also arise in non-canonical models \cite{babichev,trilogia1,trilogia2}. Even though the model is more intricate, it was shown in Refs.~\cite{trilogia1,trilogia2} a first order framework supporting kinks. Among the many classes of generalized models, there is the Born-Infeld one, which is based on the concept of limiting the electric field strength through non-linear electrodynamics. In this model, a square root appears in the dynamical term associated to the gauge field in the Lagrangian density \cite{BI1,BI2,BI3,BI4}. This allows, for instance, to regularize the electron self-energy, which is divergent in the standard Maxwell's theory. Over the years, this concept was investigated in the context of gravitation, leading to the so-called Born-Infeld gravity, in which the square root encompasses terms associated to the curvature of the spacetime \cite{bigravity1,bigravity2,bigravity3,bigravity4,bigravity5}. By extending the Born-Infeld concept to the study of scalar field models, one may find tachyonic kinks, which were first proposed in Ref.~\cite{asen1}. These objects are infinitely thin but engender finite energy. They were further investigated in Refs.~\cite{kim,huang,asen2,bazeiatach,kluson1,kluson2,kwon,roychowdhury}.

The study of thick branes was extended to non-canonical models; see, for instance, Refs.~\cite{tachyonbrane,adambrane,trilogia3,liu,twinbrane,roldao1,genbent,diazbent,roldao2,billic,roldao3,cuscuton,moazzen}. In particular, in Ref.~\cite{tachyonbrane}, the authors investigated braneworlds in which the bulk solutions are provided by tachyonic potentials. Also, in Ref.~\cite{trilogia3}, it was developed a first order framework to investigate braneworlds modeled by kinklike structures in generalized models in which the Lagrangian density is an arbitrary function of the scalar fields and its dynamical term, which is quadratic in the derivatives. In this situation, the gravity sector of the brane is stable. Moreover, in Ref.~\cite{genbent}, bent branes were investigated, through the second order differential equations, in generalized models with the presence of the 4-dimensional cosmological constant.

In this work, we study a class of Born-Infeld-like scalar field models coupled to the usual 5-dimensional gravity scenario that support flat and bent thick brane solutions. In Sec.~\ref{sec2}, we present the general model and obtain the equations which govern the brane, taking into account the null energy condition. A first order framework is developed in Sec.~\ref{sec3}, where we determine the constraints that must be solved to make it compatible with the Einstein equations. The investigation goes on in Sec.~\ref{sec4}, where we illustrate our procedure with some specific models, discussing the impact of the 4-dimensional cosmological constant. In Sec.~\ref{sec5}, we conclude the investigation with some ending comments and perspectives.

\section{Flat and bent branes}\label{sec2}

We start our investigation with the 5-dimensional Einstein-Hilbert action
\be\label{actions}
S=\int d^5x\,\sqrt{|g|}\left(-\frac14R +\LL_s\right),
\ee
in which $R$ represents the Ricci scalar, $\LL_s$ describes the source Lagrange density and $g$ denotes the 5-dimensional determinant of the metric tensor
\be\label{gab}
g_{ab}=
\begin{pmatrix}
e^{2A(y)} g^{(4)}_{\mu\nu} & 0 \\
0 & -1 
\end{pmatrix},
\ee
such that the line element is $ds_5^{2} = e^{2A(y)}ds_4^{2} - dy^2$, with $ds_4^2 = g^{(4)}_{\mu\nu} dx^\mu dx^\nu$ and $g^{(4)}_{\mu\nu}$ denoting the $4$-dimensional metric tensor. We are taking natural units and dimensionless coordinates and fields, with $4\pi G=1$. The indices $a,b$ are related to the bulk dimensions and run from $0$ to $4$, while $\mu,\nu$ are associated to the 4-dimensional spacetime, ranging from $0$ to $3$. The extra dimension is represented by the coordinate $y\in(-\infty,+\infty)$ and $A(y)$ denotes the warp function.

We then consider three possibilities for the 4-dimensional metric, with the line element being
\be\label{ds4}
ds^{2}_4 =
\begin{cases}
\displaystyle dt^2 -\left(dx_1^2 +dx_2^2 +dx_3^2\right), &\text{M}_4\\
dt^2 -e^{2\sqrt{\Lambda}t}\left(dx_1^2 +dx_2^2 +dx_3^2\right), &\text{dS}_4\,\\
e^{-2\sqrt{-\Lambda}x_3}\left(dt^2 -dx_1^2 -dx_2^2\right)-dx_3^2, &\text{AdS}_4
\end{cases}.
\ee
The parameter $\Lambda$ represents the $4$-dimensional cosmological constant. The dS$_4$ and AdS$_4$ geometries are only possible for $\Lambda>0$ and $\Lambda<0$, respectively.
 
To obtain the Einstein equation, we proceed as usual and take the variation of the action with respect to the metric tensor. This leads to
\be\label{ER}
R_{ab} -\frac12g_{ab}R = 2T_{ab},
\ee
where $R_{ab}$ and $T_{ab}$ are the Ricci and energy-momentum tensors, respectively. As we have previously commented, to obtain thick branes, one may introduce a scalar field in the action \cite{dewolfe}. We consider a Born-Infeld-like model, described by the Lagrangian density
\be\label{lmodel}
\LL_s = -V(\phi)\sqrt{1 -\n_a\phi\n^a\phi} +F(\phi),
\ee
where $V(\phi)$ is the potential and $F(\phi)$ is an arbitrary function which depends only on the scalar field. The presence of the square root imposes that $1 -\n_a\phi\n^a\phi\ge0$. Notice that this model presents the square root, similarly to the original Born-Infeld concept, but also includes the function $F(\phi)$. The energy-momentum tensor associated to the above expression is
\be\label{tab}
T_{ab} = \frac{V(\phi)\n_a\phi\n_b\phi}{\sqrt{1 -\n_c\phi\n^c\phi}} -g_{ab}\LL_s.
\ee
It allows us to investigate the null energy condition (NEC), $T_{ab}n^an^b\geq0$, where $n^a$ is a null vector that satisfies $g_{ab}n^an^b=0$. From this, we get
\be\label{nec}
\frac{V(\phi)}{\sqrt{1 -\n_a\phi\n^a\phi}} \geq 0,
\ee
which imposes that the potential must be non-negative, $V(\phi)\geq0$.

The equation of motion of the scalar field is
\be\label{eomB}
\n_a\left(\frac{V(\phi)\n^a\phi}{\sqrt{1 -\n_b\phi\n^b\phi}}\right) +V_\phi\sqrt{1 -\n_a\phi\n^a\phi} -F_\phi = 0.
\ee
By using it, one can show that the energy-momentum tensor is conserved, $\nabla^a T_{ab}=0$. The above equation may also be written in the form
\be\label{eomBexp}
\mathcal{G}^{ab}\n_a\n_b\phi + \frac{V_\phi}{\sqrt{1-\n_c\phi\n^c\phi}} -F_\phi = 0,
\ee
where
\be
\mathcal{G}^{ab} = \frac{V(\phi)}{\sqrt{1-\n_c\phi\n^c\phi}}g^{ab} +\frac{V(\phi)}{\left(1-\n_c\phi\n^c\phi\right)^{3/2}}\n^a\phi\n^b\phi.
\ee
The factor $\mathcal{G}^{ab}$ that appears in the equation of motion \eqref{eomBexp} is related to its hyperbolic character, which is ensured for $1-\n_a\phi\n^a\phi>0$ if the NEC \eqref{nec} is satisfied. Thus, for static fields in the metric described by \eqref{gab} this condition is attained.

Since the warp factor in the metric tensor \eqref{gab} depends only on the extra dimension, we consider that the scalar field $\phi$ also does it, that is, $\phi=\phi(y)$. The equation of motion \eqref{eomB} turns into
\be\label{seomB}
\frac{V(\phi)\phi^{\prime\prime}}{\left(1 +{\phi^{\prime}}^2\right)^{3/2}}+\frac{4V(\phi)A^\prime\phi^\prime-V_\phi}{\sqrt{1 +{\phi^{\prime}}^2}} +F_\phi = 0.
\ee
One can use the metric tensor \eqref{gab} and the energy-momentum tensor \eqref{tab} to show that the Einstein equation \eqref{ER} leads to
\bes\label{ERbent}
\bal
\label{ERbenta}A^{\prime\prime} +\Lambda e^{-2A} &= -\frac{2V(\phi){\phi^{\prime}}^2}{3\sqrt{1 +{\phi^{\prime}}^2}},\\
\label{ERbentb}{A^{\prime}}^2 -\Lambda e^{-2A} &= \frac13\left(F(\phi) -\frac{V(\phi)}{\sqrt{1 +{\phi^{\prime}}^2}}\right).
\eal
\ees
As expected from the conservation of $T_{ab}$, one can use these expressions to get the equation of motion \eqref{seomB}. Notice the presence of $\Lambda$ in the above equations. The Minkowski, de Sitter and anti-de Sitter spacetimes described by Eq.~\eqref{ds4} are obtained for $\Lambda=0$, $\Lambda>0$ and $\Lambda<0$, respectively.

\section{First order framework}\label{sec3}
We see that the differential equations that drive $\phi(y)$ and $A(y)$ are of second order. To simplify the problem, we develop a first order framework to investigate brane solutions. To do so, we introduce two functions which depend only on the scalar field, $G(\phi)$ and $W(\phi)$, which control the derivative of the scalar field and warp function in the form
\bes\label{fo}
\bal
\label{foG}
\phi^\prime &= G(\phi),\\
\label{foAp}
A^\prime &= -\frac13W(\phi). 
\eal
\ees
The behavior of $W$ determines which geometries are interpolated in the braneworld scenario. We then investigate the conditions under which Eqs.~\eqref{fo} are compatible with the Einstein equations \eqref{ERbent}. This requires that
\bes\label{vfbent}
\bal
V(\phi) &= \frac{W_\phi\sqrt{1 +G(\phi)^2}}{2G(\phi)} \Bigg(1 -\frac{3\Lambda h(\phi)}{W_\phi G(\phi)}\Bigg),\\ \label{fbent}
F(\phi) &= \frac{W_\phi}{2G(\phi)} +\frac{W(\phi)^2}3 -3\Lambda h(\phi)\left(1 +\frac{1}{2G(\phi)^2}\right),
\eal
\ees
where we have defined, for convenience, the function
\be\label{hint}
h(\phi) = \exp\left(\frac23\int \frac{d\phi\,W(\phi)}{G(\phi)}\right).
\ee
The above function is useful to calculate the warp function, as it can be written in the form 
\be\label{ah}
A(y) = -\frac12\ln(h(\phi)),
\ee
where $\phi=\phi(y)$ is the solution of Eq.~\eqref{foG}. We remark that an integration constant arises in $h(\phi)$, which will be chosen to obey $A(0)=0$. We then see that both $V(\phi)$ and $F(\phi)$ are determined exclusively by $G(\phi)$ and $W(\phi)$.

The parameter $\Lambda$ determines the type of the brane (see Eq.~\eqref{ds4}). In the case of flat branes, $\Lambda=0$, the functions in Eqs.~\eqref{vfbent} simplify to
\bes\label{vfflat}
\bal
\label{vflat}V(\phi) &= \frac{W_\phi\sqrt{1 +G(\phi)^2}}{2G(\phi)},\\
\label{fflat}F(\phi) &= \frac{W_\phi}{2G(\phi)} +\frac13W(\phi)^2.
\eal
\ees
We remark that the functions must be chosen obeying $W_\phi/G(\phi) \geq0$ to get a non-negative potential, as required by Eq.~\eqref{nec}. This restriction implies that $F(\phi)$ must be non-negative in this specific case. So, the solutions investigated in Refs.~\cite{asen1,asen2}, in which $F(\phi)$ is absent, will not be considered here to modeling the brane. The simplest choice allowed for $F$ is $F(\phi)=f_0$, where $f_0$ is constant. However, for $f_0<0$, the condition \eqref{nec} is violated. Also, for $f_0=0$, the potential has the form $V(\phi) = -3\sqrt{1 +G(\phi)^2}\left(2\int d\phi\,G(\phi)\right)^{-2}
$, which is negative for all $G(\phi)$ and incompatible with \eqref{nec}. Thus, we only take $f_0>0$. In this situation, Eq.~\eqref{fflat} becomes a constraint for $W(\phi)$. It supports an analytical solution for $W(\phi)$ such that the potential in Eq.~\eqref{vflat} is given explicitly in terms of $G(\phi)$. From this, we can write
\bes\label{wvf0}
\bal \label{wf0}
W(\phi) &= \sqrt{3f_0}\tanh\left(2\,\sqrt{\frac{f_0}{3}}\int d\phi\,G(\phi)\right),\\ \label{vf0}
V(\phi) &= f_0\,\sech^2\left(2\,\sqrt{\frac{f_0}{3}}\int d\phi\,G(\phi)\right)\sqrt{1 +G(\phi)^2}.
\eal
\ees
In these expressions, an integration constant arise in the process. As we will see, this specific case, with $F(\phi)=f_0$, will be considered in some examples of this paper. Notice that, contrary to the case in which $F(\phi)$ is not constant, we now do not have freedom to choose $W(\phi)$, as it is determined exclusively by $G(\phi)$.

One may wonder if $F(\phi)=f_0$ could also be investigated for $\Lambda\neq0$. Unfortunately, the constraint becomes cumbersome, so the explicit form of $W(\phi)$ and $V(\phi)$ cannot be obtained for a general $G(\phi)$ as in the above equations. Thus, this case will not be studied here. Next, we present some models of flat and bent branes using the first order equations \eqref{fo}.
\section{Examples}\label{sec4}
\subsection{Model 1}
We first take the simplest case, with
\be\label{galpha}
G(\phi) = \alpha,
\ee
where $\alpha$ is a positive parameter, for convenience, in order to get monotonically increasing scalar field solutions. The solution that comes from Eq.~\eqref{foG} is
\be\label{phireta}
\phi(y) = \alpha y,
\ee
in which we have considered $\phi(0)=0$ to fix the integration constant. This function is unbounded, ranging from $-\infty$ to $\infty$ with constant derivative.

To determine the derivative of the warp function, we use 
\be\label{wtanh}
W(\phi)=a\tanh(b\phi),
\ee
with $a$ and $b$ being positive parameters. Since $\phi$ is given as in Eq.~\eqref{phireta}, the above expression changes its sign at $y=0$. By using Eqs.~\eqref{galpha} and \eqref{wtanh}, one can show that the function in Eq.~\eqref{hint} becomes $h(\phi)= {\cosh(b\phi)}^{\frac{2a}{3\alpha b}}$. From Eqs.~\eqref{vfbent}, we have
\bes
\bal\label{pot1}
V(\phi) &= \frac{ab\sqrt{1 +\alpha^2}\,S^2}{2\alpha}\left(1 -\frac{3\Lambda}{\alpha\,a\,b\,{S}^{2+\frac{2a}{3\alpha b}}}\right),\\ \label{f1}
F(\phi) &= \frac13a^2 +\frac{a(3b-2\alpha a)\,S^2}{6\alpha} -\frac{3\Lambda (2\alpha^2+1)\,}{2\alpha^2{S}^{\frac{2a}{3\alpha b}}},
\eal
\ees
where we have defined
\be\label{sech}
S=\sech(b\phi)
\ee
for simplicity. To ensure the positiveness of the potential, as required by \eqref{nec}, solutions with positive $\Lambda$ must be discarded in this model. For $\Lambda=0$, the above potential is localized, having a bell shape with runaway minima and zeroes. For $\Lambda<0$, one has to take into account the extra term that is driven by the parameters $\Lambda$, $a$, $b$ and $\alpha$. In this regime, $\phi=0$ is a local maximum of the potential for $|\Lambda|<|\Lambda_*|\equiv \alpha^2b^2$, with two symmetric minima around it. For $|\Lambda|\geq|\Lambda_*|$, $\phi=0$ becomes a global minimum.

\begin{figure}
	\centering
	\includegraphics[width=7cm]{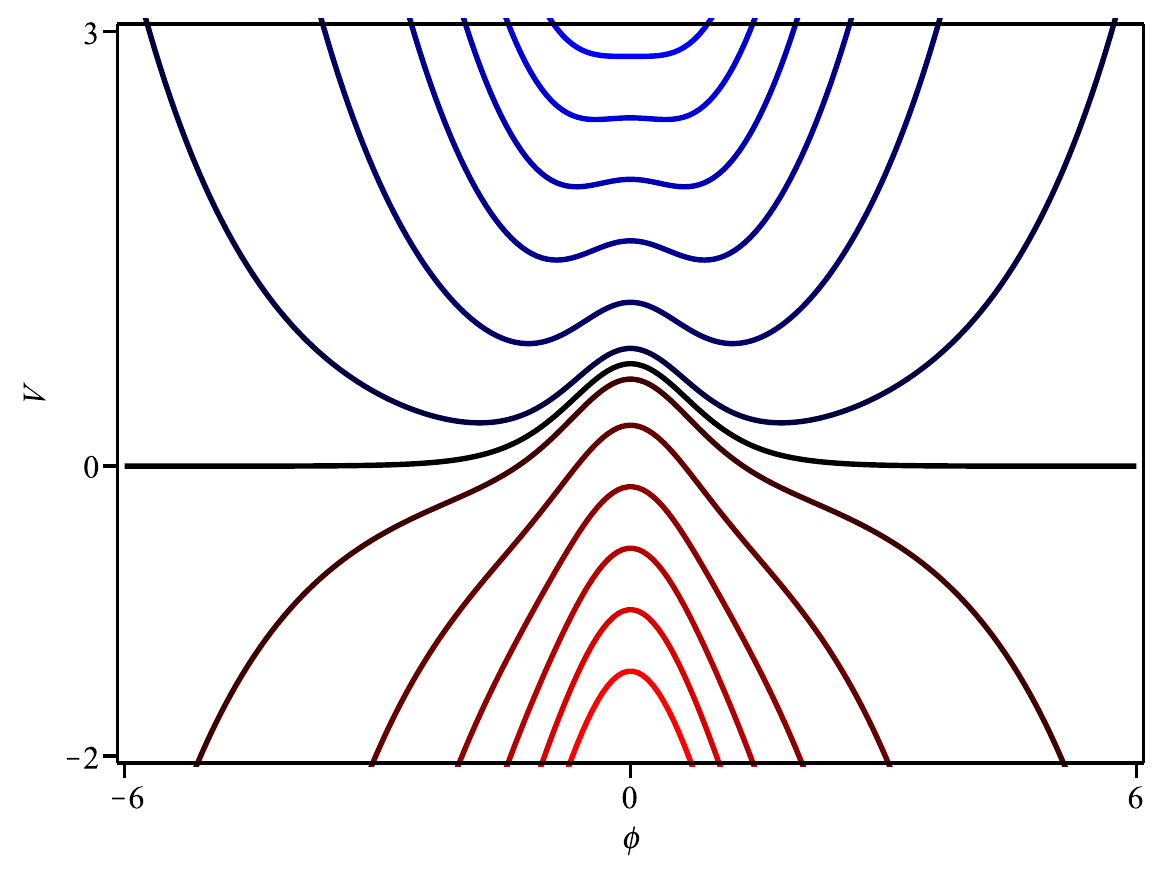}
	\includegraphics[width=7cm]{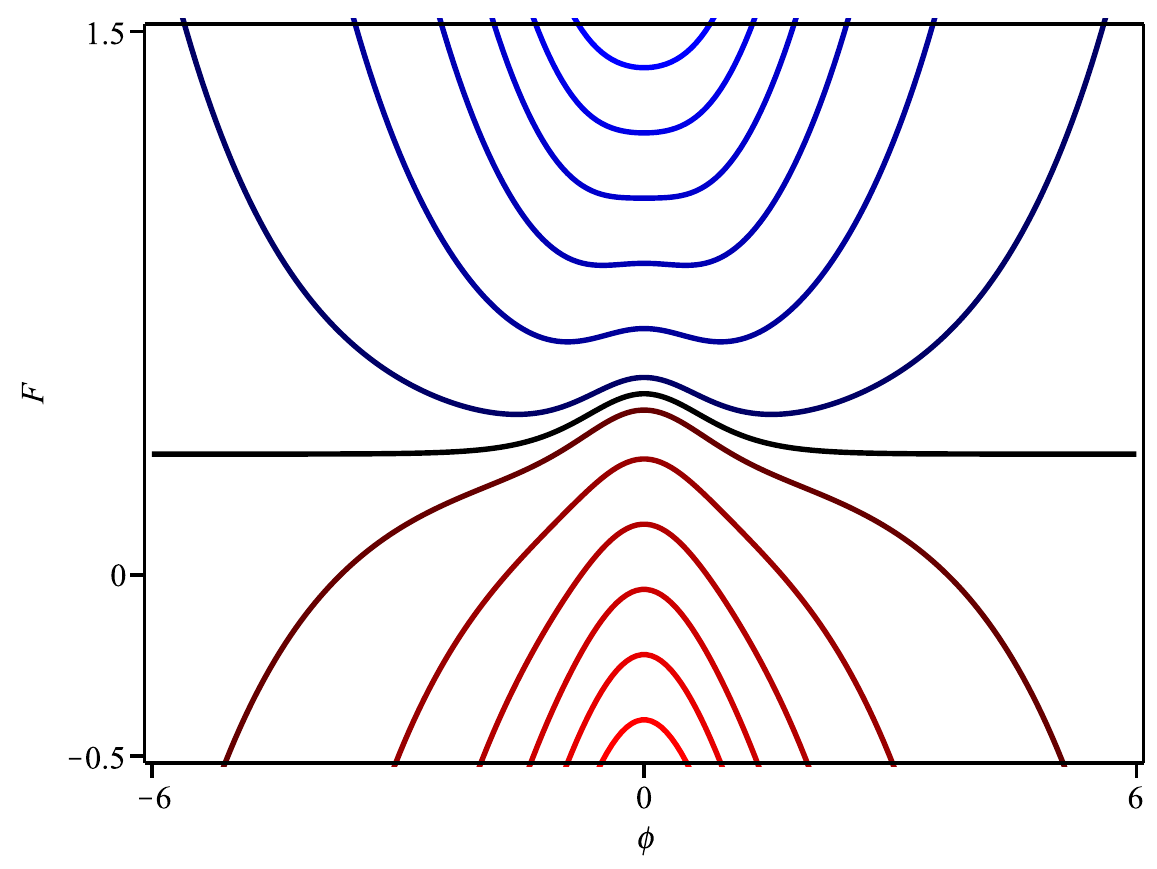}
	\caption{The potential in Eq.~\eqref{pot1} (top) and the function in Eq.~\eqref{f1} (bottom) for $\alpha=a=1$, $b=1$ and several values of $\Lambda$. For convenience, we use $\Omega=\{-1,-0.8,-0.6,-0.4,-0.2,-0.05,$ $ 0,0.05,0.2,0.4,0.6,0.8,1\}$ and take $\Lambda \in \Omega$ for the potential and $\Lambda \in \Omega/5$ for the function $F(\phi)$. The colors blue, black and red represent negative, null and positive values of $\Lambda$, respectively. The darkness of the colors increases as $\Lambda$ approaches zero.}
	\label{fig1}
\end{figure}

In Fig.~\ref{fig1}, we display the behavior of the potential for $\alpha=a=1$, $b=1/5$ and $\Lambda \in \Omega$ with $\Omega=\{-1,-0.8,-0.6,$ $-0.4,-0.2,-0.05,0,0.05,0.2,0.4,0.6,0.8,1\}$. For the function $F(\phi)$, we use $\Lambda\in\Omega/5$ to get a better visualization. The curves in blue, black and red stand for AdS, Minkowski and dS geometries, respectively. It is worth commenting that, even though the potential and the function $F(\phi)$ are defined for any value of $\Lambda$, the case $\Lambda>0$ leads to solutions that violate the NEC in Eq.~\eqref{nec}.

Compared to the potential, the function in Eq.~\eqref{f1} behaves differently with the parameters. For the flat scenario, $F(\phi)$ goes to $a^2/3$ at infinity, having a lumplike profile for $b\neq \tilde{b}$, where $\tilde{b} = 2a\alpha/3$. The lump points upwards (downwards) for $b>\tilde{b}$ ($b<\tilde{b}$). Specifically for the critical value, $b=\tilde{b}$, we have $F(\phi)=a^2/3$, which is constant and falls in the class discussed in Eqs.~\eqref{wvf0}. On the other hand, in case with the AdS$_4$ geometry, one has that $\phi=0$ defines a minimum if $b\leq\tilde{b}$; for $b>\tilde{b}$, it leads to a local maximum for $|\Lambda|<|\tilde{\Lambda}|\equiv \alpha^2 b(b-\tilde{b})/(2\alpha^2+1)$ and a global minimum for $|\Lambda|\geq|\tilde{\Lambda}|.$

The warp function can be found using Eq.~\eqref{ah}, from which one gets
\be
A(y) = {\frac{a}{3\alpha b}}\,\ln\!\left(\sech\!\left(\alpha b y\right)\right).
\ee
The behavior of this warp function is similar to the usual one for thick branes in standard dynamics (see Refs.~\cite{dewolfe,gremm1}), so we do not display it here. It leads to a warp factor, $\exp(2A(y))$, with a bell shape, having a maximum at $y=0$ and smoothly decreasing towards zero asymptotically.

We remark that, due to the simplicity of the function in Eq.~\eqref{galpha}, other models with analytical results can be found. For instance, one may take $W(\phi)=a\,\sgn(\phi)\left(1-e^{-b|\phi|}\right)$, which leads to results with features that can already be found in the model described by Eq.~\eqref{wtanh}.

\subsection{Model 2}
The scalar field is described by the function $G(\phi)$ as in Eq.~\eqref{foG}. This first order equation has the same form of the one which arises in the study of kinklike structures in $(1,1)$ dimensions. Among the many possibilities, there is the so-called vacuumless model, whose properties were well studied in Refs.~\cite{vac1,vac2}. We get inspiration from this model and consider
\be\label{G2}
G(\phi)=\alpha\,\sech(b\phi),
\ee
with $\alpha$ and $b$ being positive parameters. By substituting it in Eq.~\eqref{foG}, one gets the solution
\be\label{sol2}
\phi(y)=\frac{1}{b}\,\arcsinh(\alpha by).
\ee
As in the previous model, it is unbounded, but the derivative is not constant here. Instead, $\phi'$ vanishes asymptotically, with the tail behaving as $\phi'(y)\propto1/y$. 

\begin{figure}[t!]
	\centering
	\includegraphics[width=7cm]{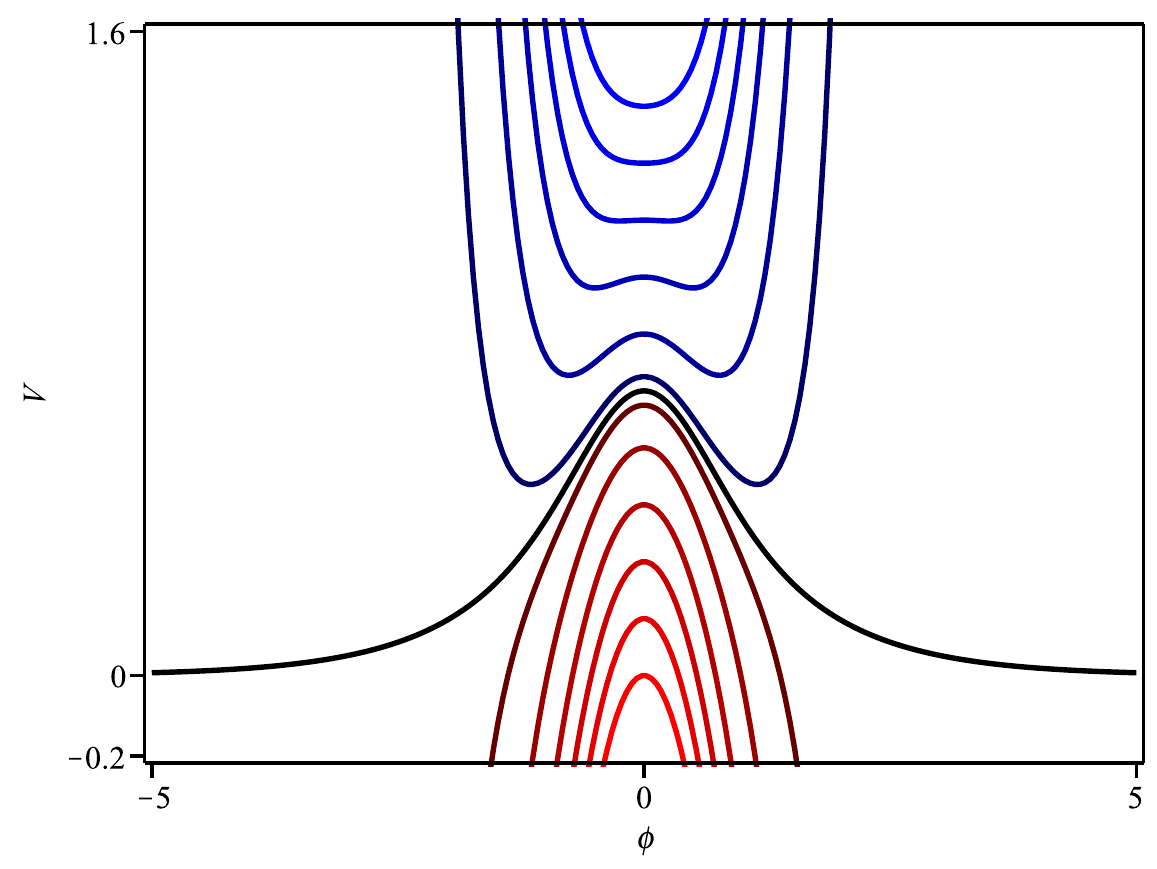}
	\includegraphics[width=7cm]{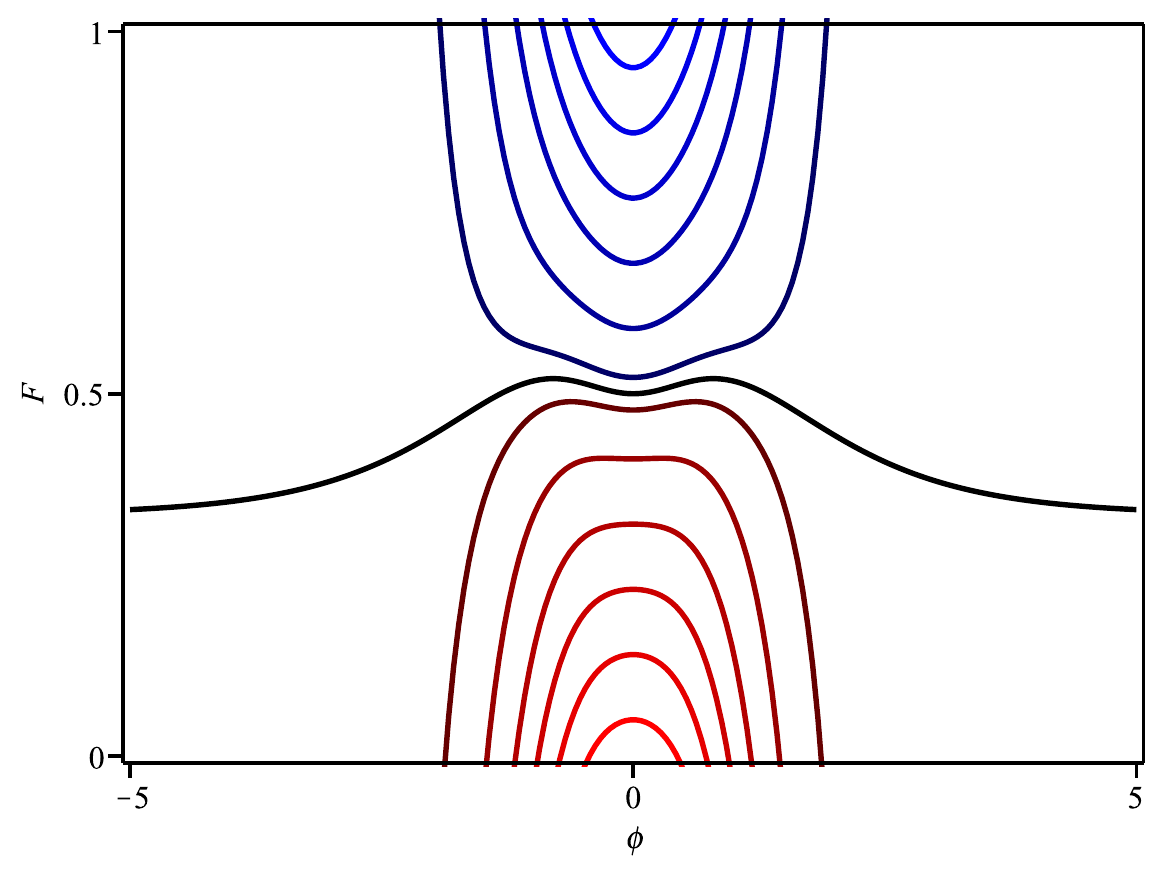}
	\caption{The potential in Eq.~\eqref{pot2} (top) and the function in Eq.~\eqref{f2} (bottom) for $\alpha=a=b=1$ and several values of $\Lambda$. We take $\Lambda \in \Omega/3$ for the potential and $\Lambda \in \Omega/10$ for the function $F(\phi)$. The colors and definition of $\Omega$ follow the previous figure.}
	\label{fig2}
\end{figure}

\begin{figure}[t!]
	\centering
\includegraphics[width=7cm]{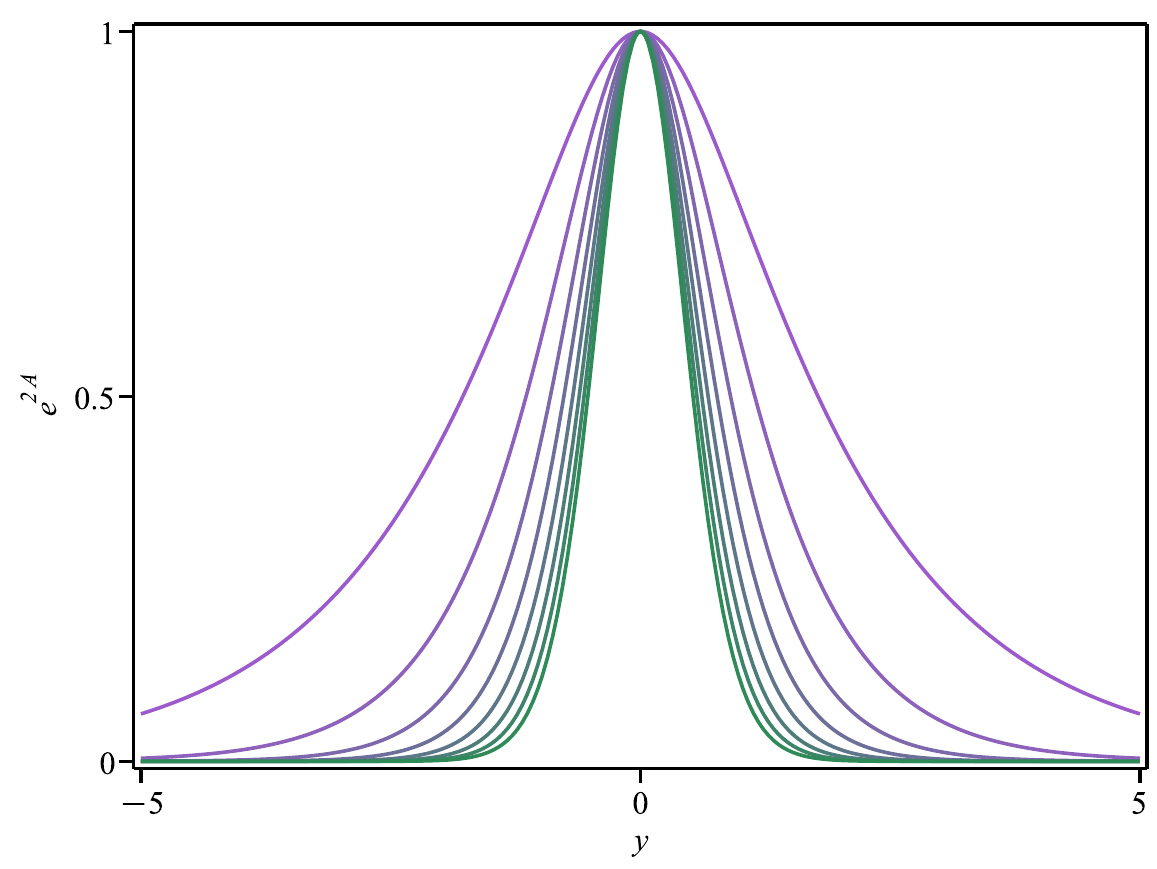}
	\caption{The warp factor, $\exp(2A(y))$, associated to the function in Eq.~\eqref{warp2} for $\alpha=b=1$ and $a=1,2,\ldots,8$. The color ranges from purple to green as one increases the parameter $a$.}
	\label{fig3}
\end{figure}

We also take the same $W(\phi)$ in Eq.~\eqref{wtanh}, which leads to $h(\phi)=\exp(2a(1-S)/(3\alpha bS))$ from Eq.~\eqref{hint}, with $S$ defined as in Eq.~\eqref{sech}. The functions $V(\phi)$ and $F(\phi)$ in Eq.~\eqref{vfbent} can be calculated analitically; they are
\bes
\bal\label{pot2}
V(\phi) &= \frac{abS\sqrt{1 +\alpha^2S^2}}{2\alpha}\left(1-\frac{3\Lambda e^{\frac{2a(1-S)}{3\alpha b S}}}{\alpha abS^3}\right),\\ \label{f2}
F(\phi) &= \frac13a^2 +\frac{abS}{2\alpha} -\frac13a^2S^2 -3\Lambda e^{\frac{2a(1-S)}{3\alpha b S}}\!\left(1 +\frac{1}{2\alpha^2S^2}\right).
\eal
\ees
Similarly to the previous model, the null energy condition in Eq.~\eqref{nec} requires that $\Lambda$ is non-positive. Considering $\Lambda=0$, one sees that the potential has the form of a bell, vanishing asymptotically and having the maximum value $V(0) = ab\sqrt{1+\alpha^2}/(2\alpha)$. In $F(\phi)$, the bell shape only appears for $3b/(4a\alpha)\geq1$, with the maximum defined by $F(0)=ab/(2\alpha)$, changing the profile for other values of $b$, as the central maximum becomes a minimum. In the case where $\Lambda<0$, both $V(\phi)$ and $F(\phi)$ diverge asymptotically. The general behavior of these functions are displayed in Fig.~\ref{fig2}.
To calculate the warp function, we use Eq.~\eqref{ah}, which leads us to
\be\label{warp2}
A(y) = \frac{a}{3\alpha b}\left(1-\sqrt{1+\alpha^2b^2y^2}\right).
\ee
The warp factor behaves near the origin as a Gaussian function, in the form $\exp(2A(y))\propto \exp(-\alpha aby^2/3)$. Far away from the origin, i.e., asymptotically, one has $\exp(2A(y))\propto\exp(-2a|y|/3)$, as expected for thick brane solutions. In Fig.~\ref{fig3}, we display the warp factor, $\exp(2A(y))$, for some values of $a$. 

We highlight that the function in Eq.~\eqref{f2} cannot be constant, regardless the values of the parameters $a$, $b$, $\alpha$ and $\Lambda$, including $\Lambda=0$. This occurs because we are choosing $W(\phi)$ as in Eq.~\eqref{wtanh}. However, we can keep $G(\phi)$ as in Eq.~\eqref{G2} and take $\Lambda=0$ to follow Eqs.~\eqref{wvf0} to determine a model in which $F(\phi)=f_0$ is constant. By doing so, we obtain
\begin{align}\label{w2const}
W(\phi) &= \sqrt{3f_0}\tanh\left(\frac{2\alpha}{b}\,\sqrt{\frac{f_0}{3}}\arctan(\sinh(b\phi))\right).\\
V(\phi) &= f_0	\,\sech^2\left(\frac{2\alpha}{b}\,\sqrt{\frac{f_0}{3}}\arctan(\sinh(b\phi))\right) \nonumber\\
	&\times\left(1 +2\alpha^2\sech(b\phi)^2\right).   
\end{align}
Although we have the explicit form of $W(\phi)$, it is more intricate, so the warp function $A(y)$ cannot be obtained analytically. We then can use the solution \eqref{sol2} combined with the function \eqref{w2const} in Eq.~\eqref{foAp} to show that the behavior of the warp factor near and far from the origin is similar to the one commented right below Eq.~\eqref{warp2}.

\subsection{Model 3}
\begin{figure}[t!]
	\centering
	\includegraphics[width=4.15cm]{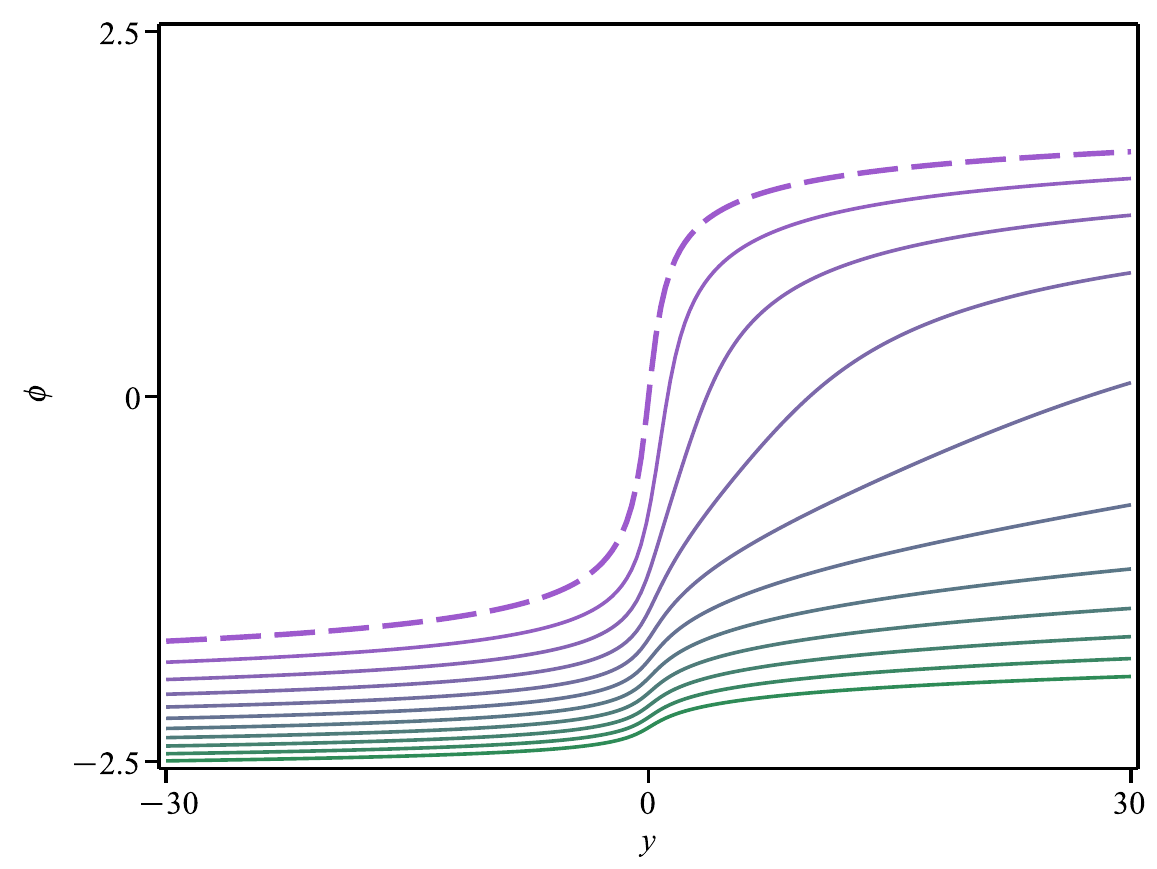}
	\includegraphics[width=4.15cm]{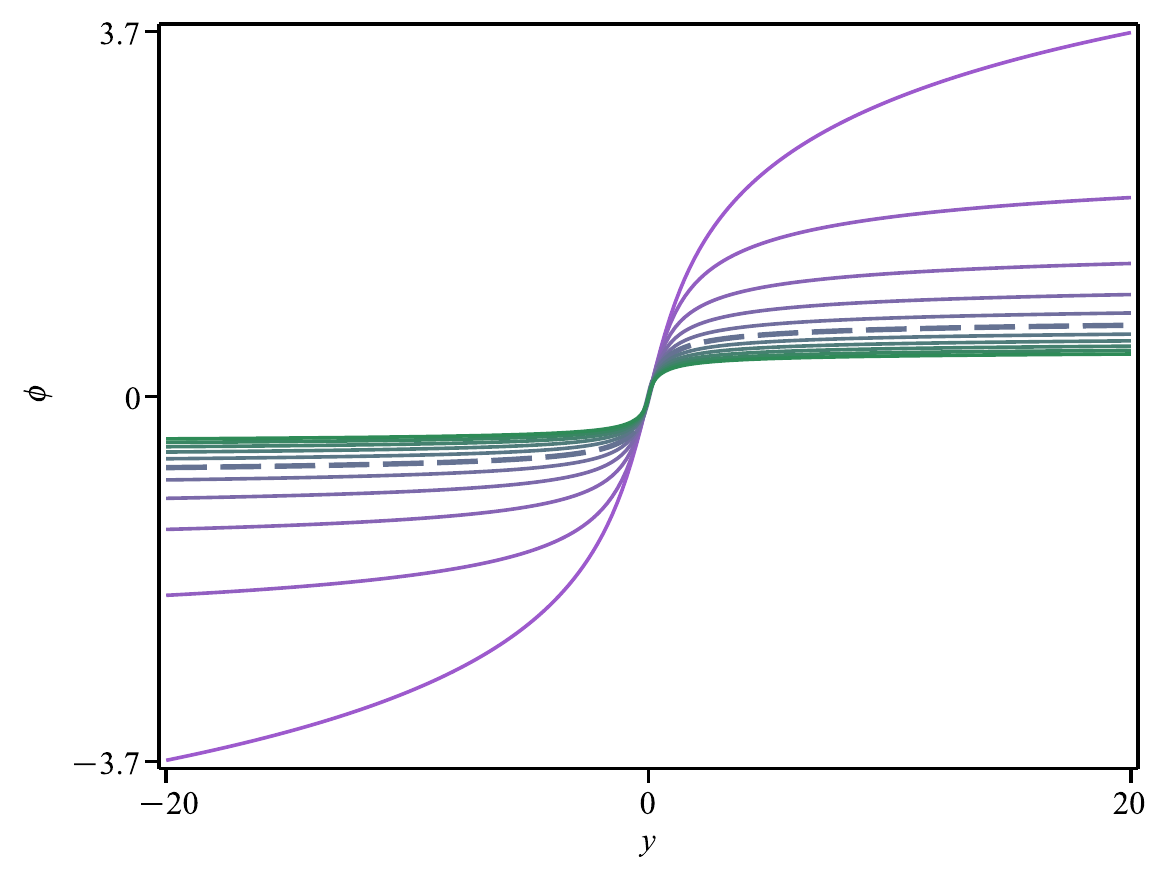}
	\includegraphics[width=4.15cm]{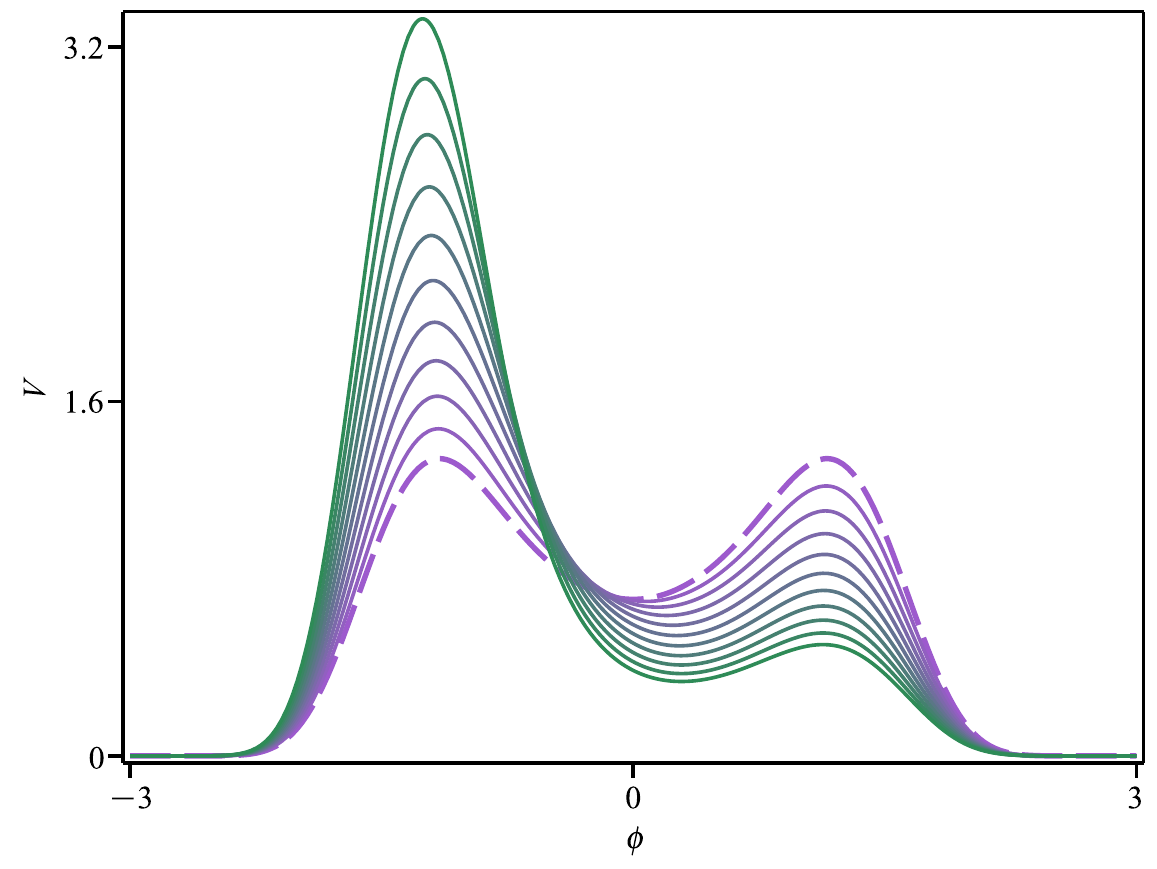}
	\includegraphics[width=4.15cm]{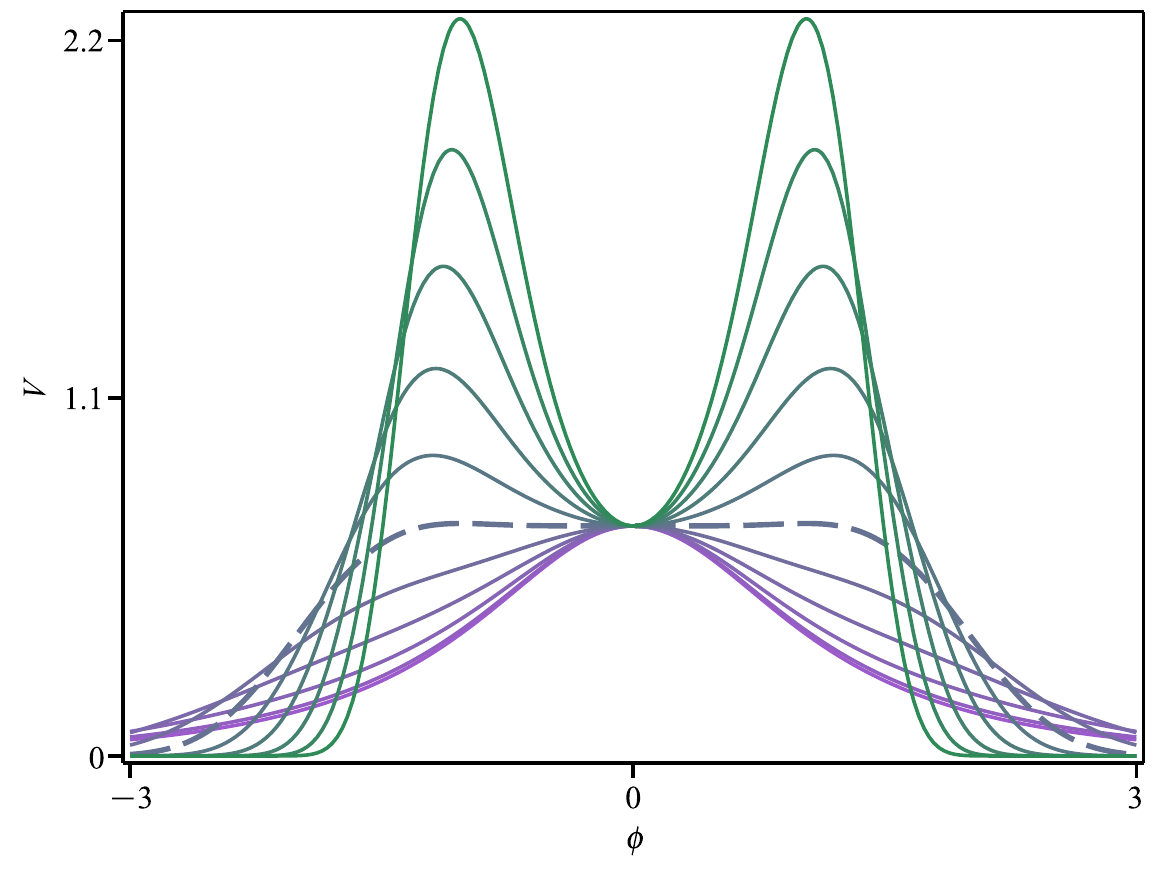}
	\includegraphics[width=4.2cm]{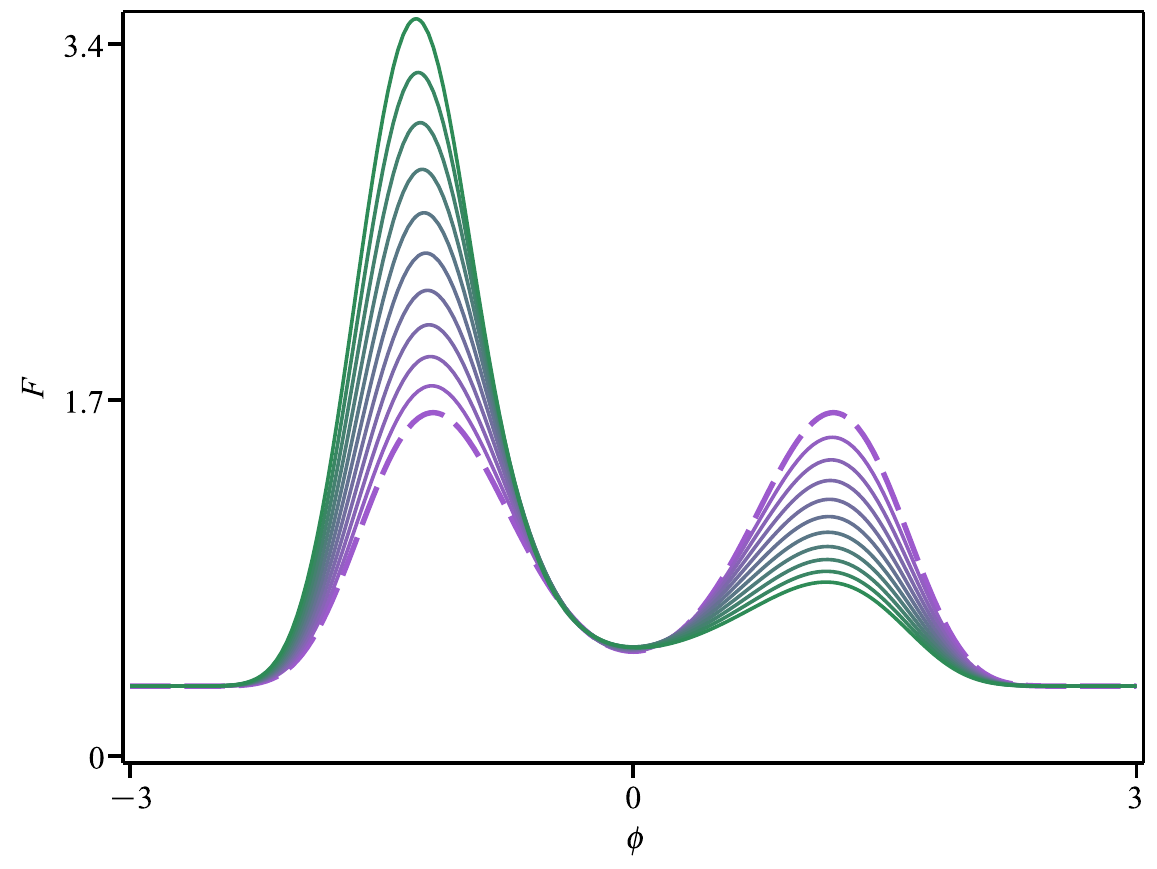}
	\includegraphics[width=4.2cm]{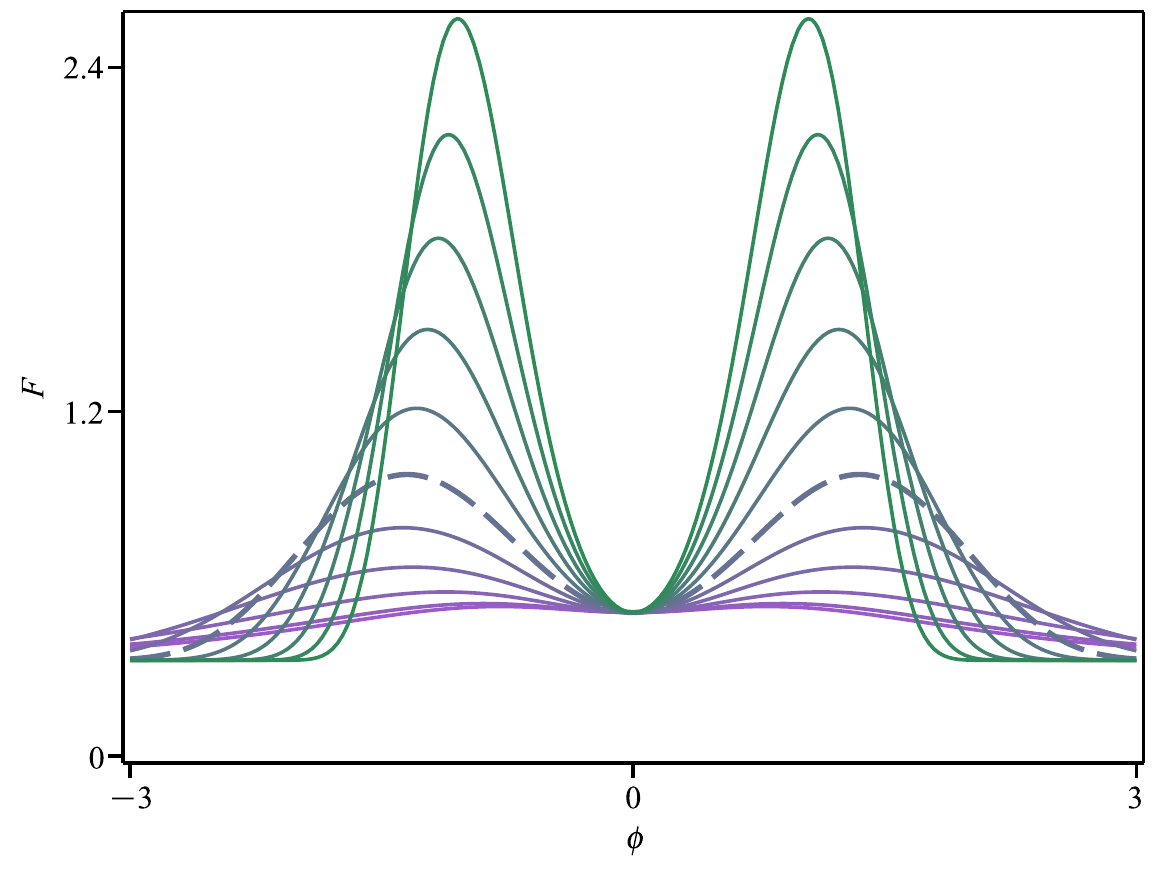}
	\caption{The solution in Eq.~\eqref{sol3} (top), potential in Eq.~\eqref{pot3} (middle) and the function in Eq.~\eqref{f3} (bottom)  for $\Lambda=0$, $\alpha=a=b=1$. The left panels are plotted for $\beta=1.5$ and $c=0,1,2,\ldots,10$ and the right ones are sketched for $c=0$ and $\beta = 0,0.2,0.4,\ldots,2$. The dashed lines stand for $c=0$ in the left panels and for $\beta=1$ in right ones. The colors represent the parameter that changes in each panel, ranging from purple (smallest value) to green (highest value).}
	\label{fig4}
\end{figure}

In the previous model, we considered the scalar field to have the vacuumless profile. We now take an extension of this model, including the presence of asymmetry in the solution, with
\be\label{G3}
G(\phi) = \alpha\,\sech(\beta\phi)\,\sech\left(cb+\frac{b}{\beta}\sinh(\beta\phi)\right),
\ee
where $c$ is a real parameter and $\alpha$, $\beta$ and $b$ are positive. By substituting this in Eq.~\eqref{foG}, we get the scalar field solution
\be\label{sol3}
\phi(y) = \frac{1}{\beta}\,\arcsinh\left(\frac{\beta}{b}\arcsinh(\alpha b y)- c\beta\right).
\ee
This solution ranges from $-\infty$ to $+\infty$, but it is asymmetric due to the presence of the parameter $c$. We display the above function in the top panels of Fig.~\ref{fig4} for some values of the parameters. In the left panel, in particular, we highlight the aforementioned feature, plotting the asymmetric ($c\neq0$, solid lines) and symmetric cases ($c=0$, dashed line). The symmetric case ($c=0$) is depicted in the right panel, where one can see that the parameter $\beta$ modifies how fast the solution goes to the infinities, while keeps the derivative constant at its center, with $\phi^\prime(0)=\alpha$. 

\begin{figure}[t!]
	\centering
	\includegraphics[width=7cm]{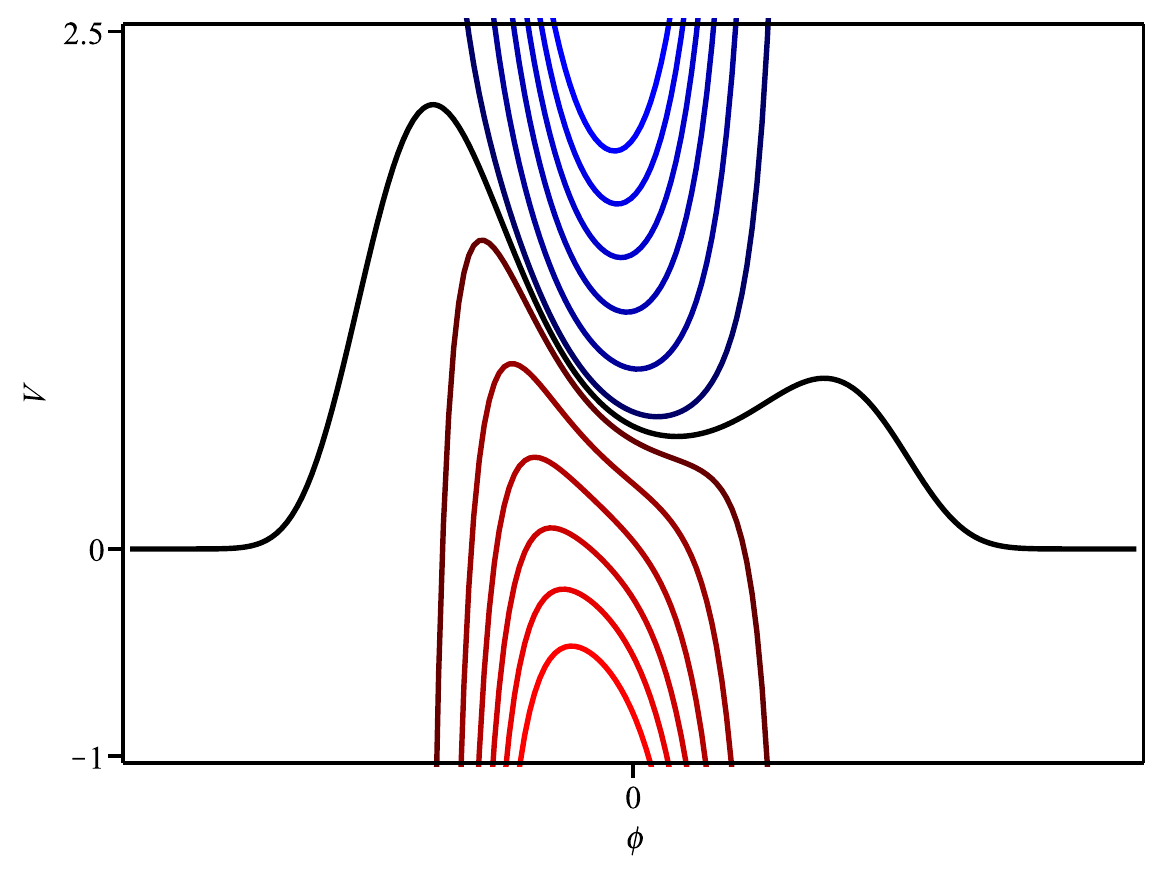}
	\includegraphics[width=7cm]{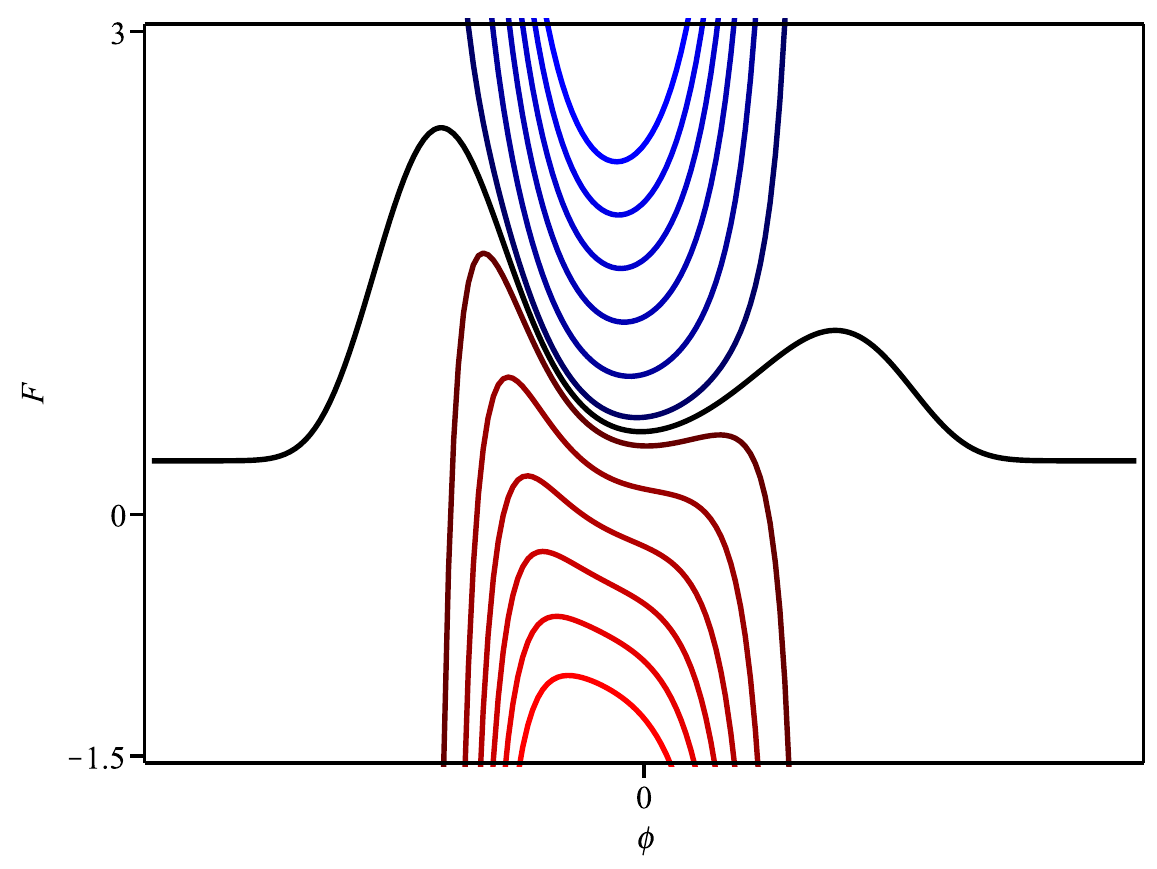}
	\caption{The potential in Eq.~\eqref{pot3} (top) and the function in Eq.~\eqref{f3} (bottom) for $\alpha=a=b=1$, $c=0.5$, $\beta=1.5$ and several values of $\Lambda$. We take $\Lambda \in \Omega/2$ for the potential and $\Lambda \in \Omega/3$ for the function $F(\phi)$. The colors and definition of $\Omega$ follow Figs.~\ref{fig1} and \ref{fig2}.}
	\label{fig5}
\end{figure}

From Eq.~\eqref{foAp}, we see that one has to provide the function $W(\phi)$ to get the warp function. Since we are dealing with the scalar field solution \eqref{sol3}, we consider a generalization of the function in Eq.~\eqref{wtanh} to get the very same warp function in Eq.~\eqref{warp2}. It is given by
\be
W(\phi) = a\tanh\left(cb+\frac{b}{\beta}\sinh(\beta\phi)\right).
\ee
Notice that Eq.~\eqref{wtanh} is recovered for $\beta\to 0$ and $c=0$. This regime of parameters also leads to $G(\phi)$ and $W(\phi)$ as in Eqs.~\eqref{G2} and \eqref{wtanh}. An interesting feature supported by the above function is that it leads to a symmetric warp function (see Fig.~\ref{fig3}), even though the scalar field solution can be asymmetric. Albeit we are dealing with functions $G(\phi)$ and $W(\phi)$ more intricate than the previous ones, we were able to obtain an analytical expression for $h(\phi)$, in the form
\be
h(\phi)= \exp\left(\frac{2a}{3\alpha b}\left( \cosh\left(cb+\frac{b}{\beta}\sinh(\beta\phi)\right)-1\right)\right).
\ee
The potential and $F(\phi)$ are
\bes
\bal\label{pot3}
V(\phi) &= \frac{abS_b\sqrt{1 +\alpha^2S_\beta^2S_b^2}}{2\alpha S_\beta^2}\left(1 -\frac{3\Lambda e^{\frac{2a(1-S_b)}{3\alpha bS_b}}}{\alpha ab S_b^3}\right),\\ \label{f3}
F(\phi) &= \frac13a^2 +\frac{a}{6\alpha}\left(\frac{3b}{S_\beta^2} -2\alpha aS_b\right)S_b\nn
	&-3\Lambda e^{\frac{2a(1-S_b)}{3\alpha b S_b}}\!\left(1 +\frac{1}{2\alpha^2S_\beta^2S_b^2}\right),
\eal
\ees
where
\be
S_b=\sech\!\left(cb+\frac{b}{\beta}\sinh(\beta\phi)\right) \quad\text{and}\quad S_\beta=\sech(\beta\phi).
\ee
The parameters $\beta$ and $c$ lead to new features, as they can induce asymmetry in $V(\phi)$ and $F(\phi)$ for $c\neq0$ and $\Lambda\leq0$. We highlight that, in the case of flat branes with $c=0$, the potential \eqref{pot3} does not present the bell-shaped form for $\beta>b\sqrt{(2\alpha^2+1)/(\alpha^2+2)}$, in contrast to the previous models that always support bell-shaped functions for $\Lambda=0$. In Fig.~\ref{fig4} we display $V(\phi)$ and $F(\phi)$ for some values of the parameters in the case $\Lambda=0$. The behavior of these functions in terms of $\Lambda$ is shown in Fig.~\ref{fig5}.

\subsection{Model 4}
\begin{figure}[t!]
	\centering
	\includegraphics[width=7cm]{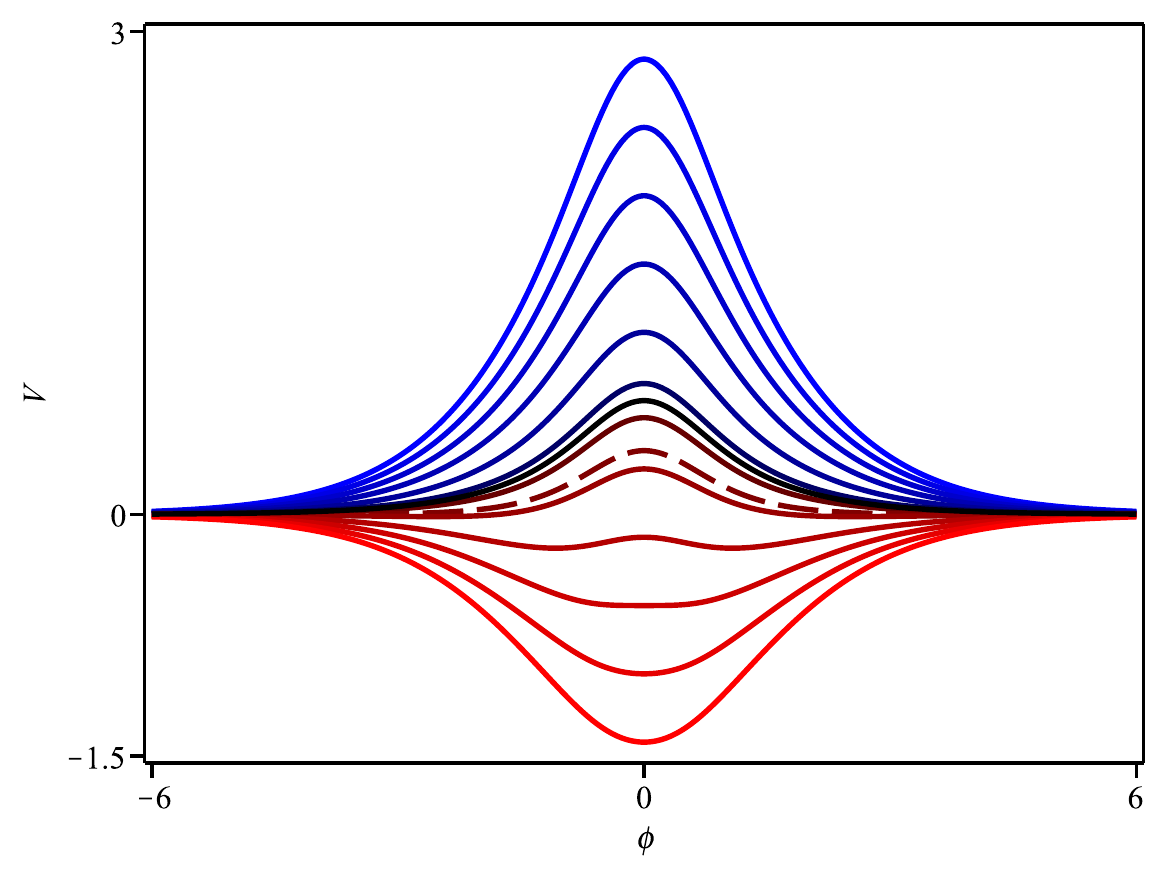}
	\includegraphics[width=7cm]{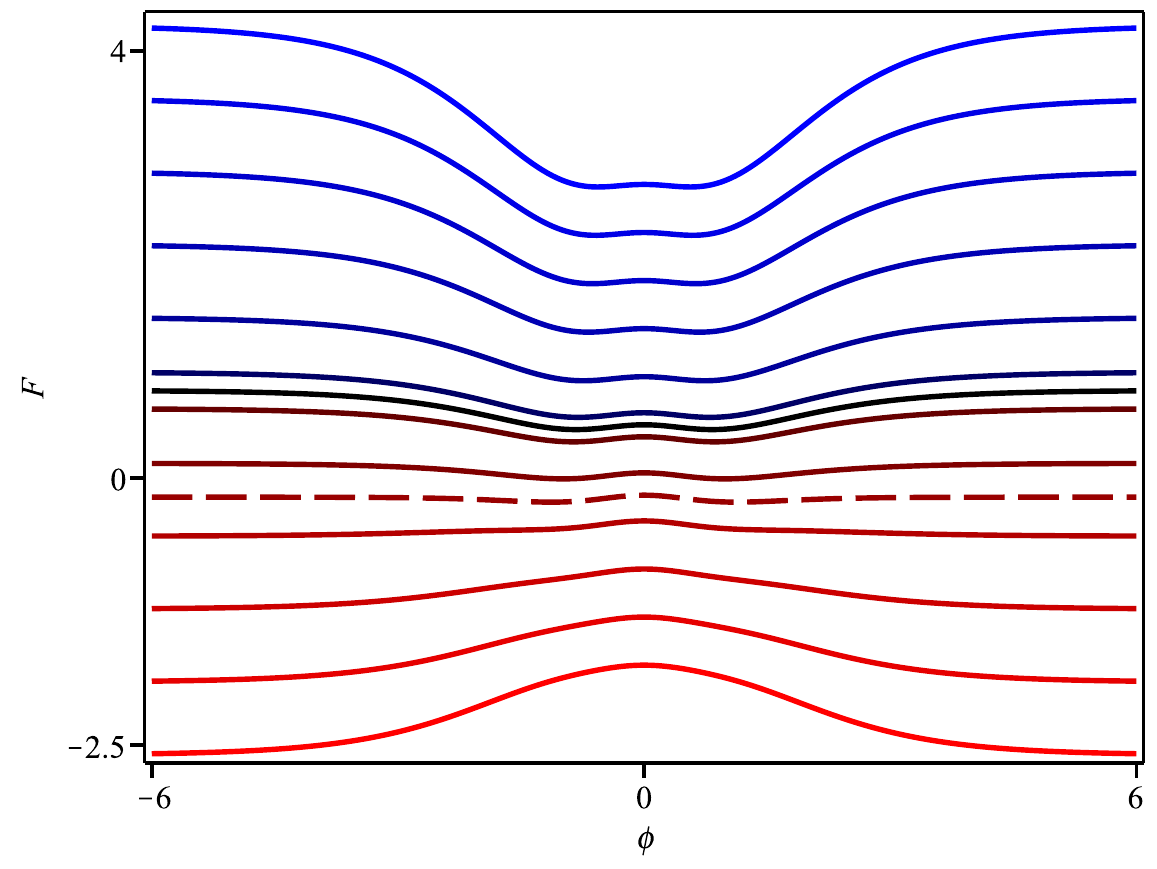}
	\caption{The potential in Eq.~\eqref{pot4} (top) and the function in Eq.~\eqref{f4} (bottom) for $\alpha=a=b=1$ and several values of $\Lambda$. We take $\Lambda \in \Omega$ for the potential and $\Lambda \in \Omega/2$ for the function $F(\phi)$. The colors and definition of $\Omega$ follow the previous figure. Also, we include the value ${\bar{\Lambda}}$ defined in Eq.~\eqref{lambc}, which is represented by the dashed lines.}
	\label{fig6}
\end{figure}

The previous model presented a procedure to introduce asymmetry in AdS$_4$ geometries (see Eq.~\eqref{ds4}). Next, we unveil the possibility of obtaining solutions for dS$_4$ spacetimes, in addition to M$_4$ and AdS$_4$ ones. To do so, we consider
\be\label{G4}
G(\phi)=\alpha\cosh(b\phi).
\ee
As before, $\alpha$ and $b$ are positive parameters. From Eq.~\eqref{foG}, one gets the solution
\be\label{sol4}
\phi(y) =
\begin{cases}
	\cfrac{2}{b}\,\arctanh\left(\tan\left(\cfrac{\alpha by}{2}\right)\right),& |y|<y_c\\
	\text{sgn}(y)\,\infty,& |y|>y_c 
\end{cases},
\ee
where $y_c=\pi/(2\alpha b)$. Notice that the above solution has infinite range, but it goes from $-\infty$ to $+\infty$ in a limited space. To calculate the warp function, we take
\be
W(\phi)=a\,\arctan(\sinh(b\phi)).
\ee
By combining it with Eq.~\eqref{hint}, we get $h(\phi)= \exp((a/3\alpha b)\arctan(\sinh(b\phi))^2)$. In this case, from Eq.~\eqref{vfbent}, we get
\bes
\bal\label{pot4}
V(\phi) &= \frac{abS\sqrt{\alpha^2 +S^2}}{2\alpha} \\ \nonumber
&\times\left( 1 -\frac{3\Lambda}{\alpha ab}\,\exp\left({\frac{a\arctan(\sinh(b\phi))^2}{3\alpha b}}\right)\right), \\ \label{f4}
F(\phi) &= \frac{abS^2}{2\alpha} +\frac{a^2\arctan(\sinh(b\phi))^2}{3}\\
	&-\frac{3\Lambda(2\alpha^2+S^2)}{2\alpha^2}\,\exp\left({\frac{a\arctan(\sinh(b\phi))^2}{3\alpha b}}\right). \nonumber
\eal
\ees
with $S$ defined as in Eq.~\eqref{sech}. The NEC in Eq.~\eqref{nec} imposes that $\Lambda$ must be in the interval $\Lambda\in(-\infty,\bar{\Lambda}]$, in which
\be\label{lambc}
\bar{\Lambda} = \frac{\alpha ab}{3e^{\frac{a\pi^2}{12\alpha b}}}.
\ee
Since the parameters $\alpha$, $a$ and $b$ are positive, the possibility of dS$_4$ geometries is included here. Thus, this model supports the three cases described in Eq.~\eqref{ds4}. Differently from the previous models, the potential \eqref{pot4} becomes null for $\phi\to\pm\infty$ for all $\Lambda$. The functions $V(\phi)$ and $F(\phi)$ are shown in Fig.~\ref{fig6} for several values of $\Lambda$. Notice that the allowed range of $\Lambda$ goes from the region in blue until the one in red, delimited by the dashed lines, which represent the value $\bar{\Lambda}$, as commented right above Eq.~\eqref{lambc}.

\begin{figure}[htb!]
	\centering
	\includegraphics[width=7cm]{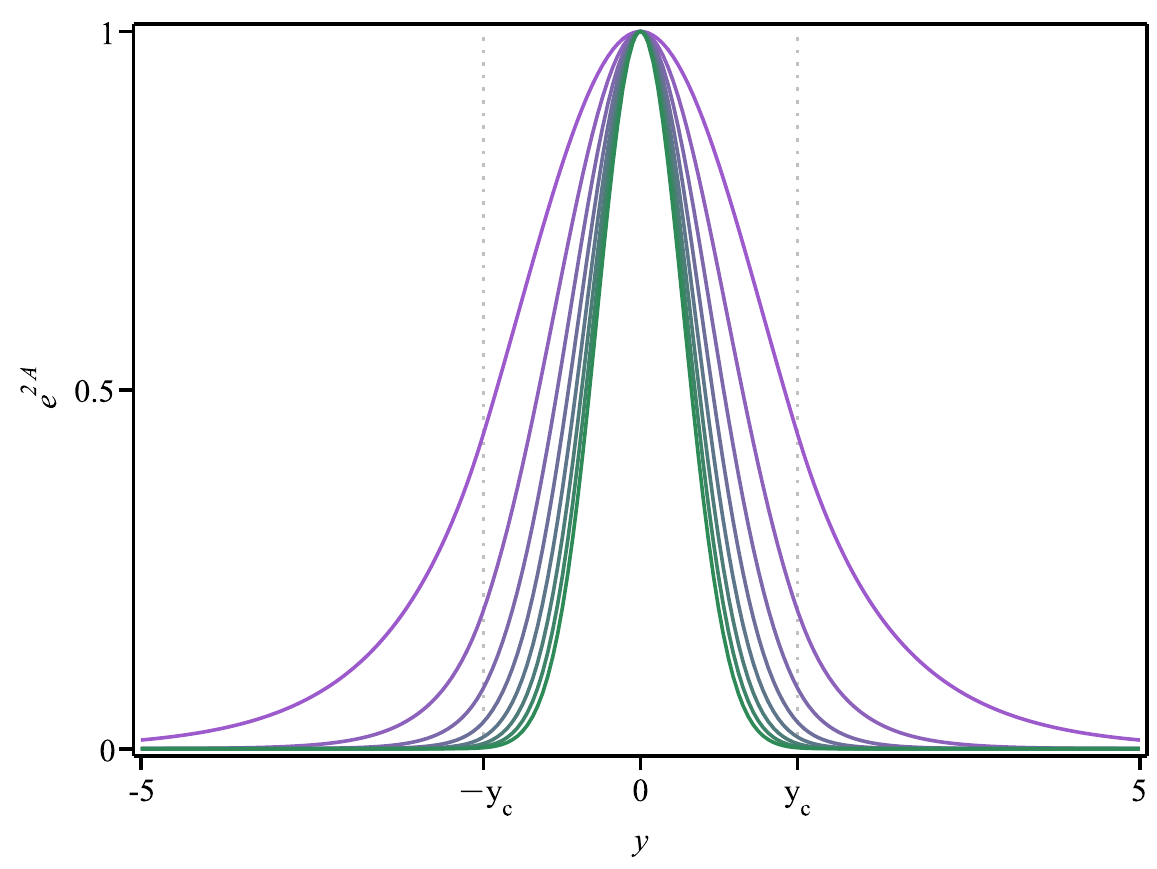}
	\caption{The warp factor $\exp(2A(y))$ associated to the function in Eq.~\eqref{warp4} for $\alpha=b=1$ and $a=1,2,\ldots,8$. The vertical lines in gray delimit the compact space $[-y_c,y_c]$, with $y_c=\pi/2$.}
	\label{fig7}
\end{figure}

The warp function is obtained directly from Eq.~\eqref{ah}, which leads to
\be\label{warp4}
A(y) = \begin{cases}
-\cfrac{\alpha ab}{6}\,y^2,& |y|\leq y_c\\
-\cfrac{a\pi}{6}\,|y|+\cfrac{a\pi^2}{24\alpha b},& |y|>y_c.
\end{cases}
\ee
The warp factor, $\exp(2A)$, associated to the above function, has an hybrid behavior. It engenders a thick (thin) brane profile whether one is inside (outside) the compact space delimited by $y_c$. Due to this feature this type of brane is called \emph{hybrid} brane \cite{t13,t14}. In Fig.~\ref{fig7}, we display the warp factor associated to the above function for some values of $a$.

\section{Final Remarks}\label{sec5}
In this work, we have investigated braneworlds modeled by scalar fields described by Born-Infeld-like dynamics as in Eqs.~\eqref{actions} and \eqref{lmodel}. We have considered the geometry associated to the 4-dimensional metric as M$_4$, dS$_4$ or AdS$_4$, in the form \eqref{ds4}. We then imposed the NEC to show that potential $V(\phi)$ must be non-negative. Since the equations that governs the brane are of second order, we have introduced the functions $G(\phi)$ and $W(\phi)$ to develop a first order framework. We remark that the presence of $F(\phi)$ allows us to have freedom in to choose $G(\phi)$ and $W(\phi)$ independently. We have shown that, although one may consider the case of constant or null $F(\phi)$, this freedom goes away, constraining both $V(\phi)$ and $W(\phi)$ to depend only on $G(\phi)$.

To illustrate our procedure, we provided four examples. In the first one, we have taken the scalar field to be unbounded with constant derivative and the resulting warp function to be analytical. In the second model, the scalar field has infinite range with the derivative vanishing asymptotically, but the warp factor behaves as a Gaussian function near the origin and falls off as a thin brane asymptotically, as usual. The third model was constructed to engender asymmetric scalar field, but symmetric warp function. In the three aforementioned models, $\Lambda$ cannot be positive due to the NEC. To unveil the possibility of the existence of potentials obeying the NEC for positive $\Lambda$, we have presented the fourth model. In this situation, $\Lambda$ has an upper bound defined by the value in Eq.~\eqref{lambc}. Interestingly, the scalar field solution goes from $-\infty$ to $+\infty$ in a limited space and the brane engender a hybrid character, behaving as a thick brane for $y\in[-y_c,y_c]$ or thin brane for $y\notin[-y_c,y_c]$, with $y_c$ defined as below Eq.~\eqref{sol4}.

As perspectives for future investigations we have, for instance, the study of Born-Infeld models in curved spacetime in different contexts, such as the study of brane inflation \cite{binfl} and black holes \cite{bh}. One may also investigate the localization of fermions on branes with the scalar field Lagrangian density in Eq.~\eqref{lmodel} in the lines of Ref.~\cite{dutrahott}. As a last perspective that we can suggest, one may study the model proposed in the current paper with the Palatini approach to modified theories of gravity \cite{palatini1,palatini2}.

\acknowledgements{We would like to thank D. Bazeia for the discussions that have contributed to this work. We acknowledge financial support from the Brazilian agencies Conselho Nacional de Desenvolvimento Cient\'ifico e Tecnol\'ogico (CNPq), grants 306151/2022-7 (MAM) and 310994/2021-7 (RM), and Paraiba State Research Foundation (FAPESQ-PB) grants 0003/2019 (RM) and 0015/2019 (MAM).}

\end{document}